\documentclass{jfm}
\usepackage{graphicx}
\usepackage{epstopdf, epsfig}
\usepackage{bm}
\usepackage{amsmath}
\usepackage{overpic}
\usepackage{float}
\usepackage{color,contour}
\usepackage{amssymb}
\usepackage[dvipsnames]{xcolor}
\usepackage{tabularx}
\usepackage{booktabs}
\usepackage{graphicx}
\usepackage{soul}
\usepackage{subfigure}
\usepackage{cite}

\usepackage{siunitx}
\usepackage[colorlinks,citecolor = blue, linkcolor= blue,hyperindex,CJKbookmarks]{hyperref}

\shorttitle{Global and local statistics in turbulent emulsions}
\shortauthor{L. Yi, F. Toschi and C. Sun}

\title{Global and local statistics in turbulent emulsions}

\author{
	Lei Yi,\aff{1}
	Federico Toschi,\aff{2,3}
	\and 
	Chao Sun\aff{1,4,5}
	\corresp{\email{chaosun@tsinghua.edu.cn}}
}

\affiliation{
	\aff{1}Center for Combustion Energy, Key Laboratory for Thermal Science and Power Engineering of Ministry of Education, Department of Energy and Power Engineering, Tsinghua University, 100084 Beijing, China

	\aff{2}Department of Physics and Department of Mathematics and
	Computer Science, Eindhoven University of Technology, 5600 MB
	Eindhoven, The Netherlands
	
	\aff{3}CNR-IAC, Via dei Taurini 19, 00185 Roma, Italy
	
	\aff{4}Department of Engineering Mechanics,
	School of Aerospace Engineering, Tsinghua University, Beijing
	100084, China

	\aff{5}Physics of Fluids Group, Max Planck-University of Twente Centre for Complex Fluid Dynamics, University of Twente, 7500 AE Enschede, The Netherlands
	
}

\date{\today}

\begin{document}

\maketitle

\begin{abstract}
	Turbulent emulsions are complex physical systems characterized by a
	strong and dynamical coupling between small-scale droplets and
	large-scale rheology. By using a specifically designed Taylor-Couette (TC) shear flow system, we are able to characterize the statistical properties of a turbulent emulsion made
	of oil droplets dispersed in an ethanol-water continuous solution, at
	the oil volume fraction up to 40\%. We find that
	the dependence of the droplet size on the Reynolds number of the flow at the volume fraction of 1\% can
	be well described by Hinze's criterion. The
	distribution of droplet sizes is found to follow a log-normal
	distribution, hinting at a fragmentation process as the possible
	mechanism dominating droplet formation.
	Additionally, the effective viscosity of the turbulent emulsion increases
	with the volume fraction of the dispersed oil phase, and decreases when
	the shear strength is increased. We find that the dependence of the
	effective viscosity on the shear rate can be described by the Herschel-Bulkley model, with a flow index
	monotonically decreasing with increasing the oil volume fraction. This
	finding indicates that the degree of shear thinning systematically
	increases with the volume fraction of the dispersed phase. 
	The current findings have important implications
	for bridging the knowledge on turbulence and low-Reynolds-number emulsion flows to turbulent emulsion flows.
\end{abstract}

\section{Introduction}

Emulsions consisting of two immiscible liquids, such as oil and water mixtures,
are omnipresent in many industrial processes, including chemical
engineering
\citep{wang2007oil}, food
processing \citep{mcclements2007critical}, drug
delivery systems
\citep{spernath2006microemulsions}, and
enhanced oil recovery
\citep{kilpatrick2012water,mandal2010characterization},
among others. While the applications of emulsions are wide, as mentioned above, the understanding of the physics of emulsions, particularly turbulent emulsions, is still rather limited.

In very low volume fraction regimes, turbulent emulsions are mainly characterized by the breakup of droplets, and coalescence events can be neglected due to the very slight chance of coalescing. The microscopic droplet structure (droplet size distribution) is generated by the turbulent stresses, while it has little influence on the macroscopic properties (viscosity) of the fluid. \cite{pacek1998sauter,pacek1994structure} conducted experimental studies that focused on turbulent emulsions in a stirred
vessel and found that the dispersed droplet size follows a log-normal
distribution. The dispersed droplet size of the emulsion in dilute regimes in a homogeneous and isotropic turbulent flow was initially investigated by
\cite{hinze1955fundamentals}, who linked the turbulent fluctuations
to the breakup of dispersed droplets, and derived an expression for the
maximum droplet size for a given intensity (i.e., Reynolds number) of a homogeneous and isotropic turbulent flow. More recently, a fully resolved numerical investigation of the droplet size distribution in homogeneous isotropic turbulence also supported the validity of the Hinze relation on the average droplet size in turbulence \citep{perlekar2012droplet}.
\textcolor{black}{
	Droplet size distribution for liquid-liquid emulsions in Taylor-Couette flows was studied based on the Kolmogorov turbulence theory~\citep{farzad2018investigation}. \cite{lemenand2017turbulent} investigated the drop size distribution in an inhomogeneous turbulent flow using a turbulent spectrum model for drop-breakup mechanisms.
}

With the increase of the volume fractions of the dispersed phase, the turbulent emulsions are characterized by the interplay between droplets breakup and coalescence events. Droplet shapes and sizes respond to and influence the macroscopic flow properties. The effective viscosity is a primary parameter among these properties. One important factor that affects the viscosity of emulsions is the volume fraction of the dispersed phase. However, the current viscosity-concentration relations for emulsions are mainly based on the analogy with that of suspensions with solid spheres \citep{pal1992rheology,derkach2009rheology}.
Many empirical equations were proposed to describe the effective viscosity of the solid particle suspension as a function of the volume fraction, such as the one proposed by Krieger and Dougherty that works for particle-fluid suspensions in both low and high concentration limits
\citep{krieger1959mechanism,krieger1972rheology}:
$\eta_{r}=(1-\phi/\phi_{m})^{-2.5\phi_{m}}$, where $\phi$ denotes the volume fraction of the solid spheres in the suspension. In this equation, the maximum volume fraction $\phi_{m}$, where the viscosity of the suspension diverges, is introduced. 
However, there are some key differences between turbulent emulsions and the suspension systems with particles. In these suspension fluids, a microscopic structure is always present and the flow can only interact with it. In fluids emulsions, however, the microscopic droplet structure, that confers the complex rheological properties to the fluid, is itself generated by the macroscopic (turbulent) stress through deformation, break-up, and coalescence of the droplets.
\textcolor{black}{
	Another empirical equation to describe the effective viscosity of the suspension is the Eilers formula, $\eta_{r}=[1+B\phi/(1-\phi/\phi_{m})]^2$, which fits well both the experimental and numerical data~\citep{stickel2005fluid,singh2003experimental,zarraga2000characterization}. In this expression, $B$ is a constant, and $\phi_{m}$ is the geometrical maximum packing fraction. Numerical studies of \cite{rosti2018rheology} show that Eilers formula is a good description also for suspensions of viscoelastic spheres, provided that the volume fraction $\phi$ is replaced by the effective volume fraction. 
	Among the studies on the effective viscosity, various types of dispersed entities have been investigated, such as deformable particles in suspensions and droplets in emulsions
	\citep{tadros1994fundamental,adams2004influence,saiki2007effects,faroughi2015generalized,rosti2018rheology,villone2019dynamics,derkach2009rheology,de2019effect}. The conventional way to measure the viscosity of a fluid is usually based on capillary tubes or rheometers, which both only operate in the laminar regime~\citep{pal1992rheology}. To determine the effective viscosity of emulsion under flowing conditions, the most usual way is to measure the pressure drop in a pipe when the emulsion flows through. \cite{urdahl1997viscosity} performed viscosity measurements of water-in-crude-oil emulsions under flowing conditions using a high-pressure test wheel.
}
The vast majority of work on emulsions focused on relatively low Reynolds numbers flows. The current knowledge of the detailed interplay between the dispersed droplets and the global rheological properties of the droplet-liquid emulsions under turbulent flow conditions is still limited.

In this work, we aim to study the emulsion in a turbulent shear flow, focusing on two aspects: (i) the statistical properties of the dispersed droplets for different Reynolds numbers at a low volume fraction; (ii) the global rheological properties of the emulsions, particularly, at high volume fractions.

\section{Experimental setup and procedure} 

\begin{figure*}
	\centering
	\includegraphics[width = 1\textwidth]{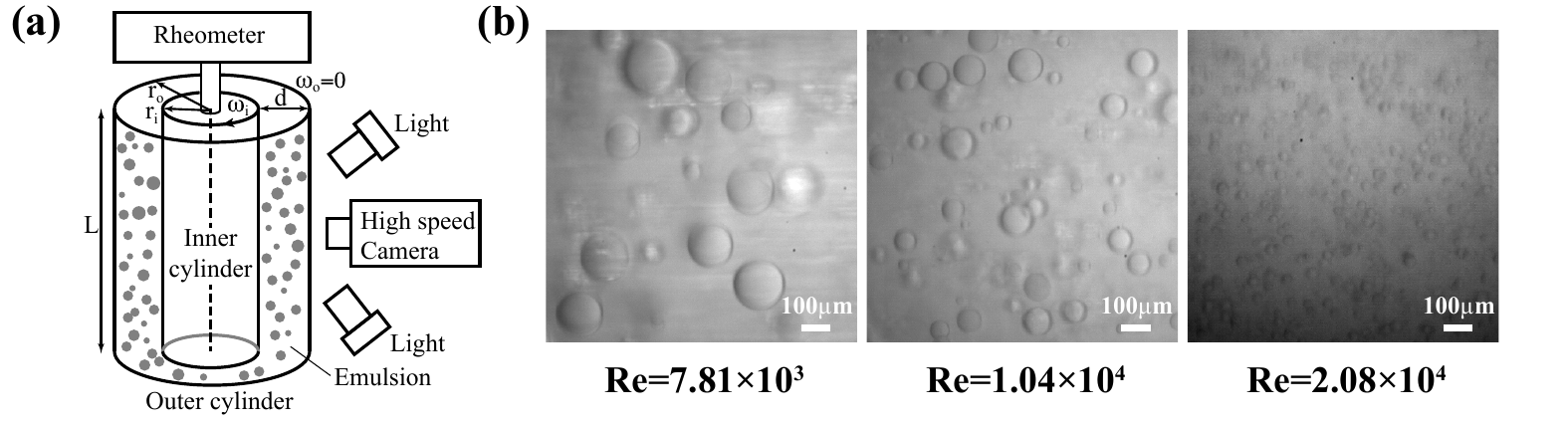}
	\caption{
		(a) The sketch of the experimental set-up. The gap between the
		outer and the inner cylinders is filled with the emulsion. The inner cylinder is connected to a rheometer so that the torque of
		the inner cylinder is directly measured by the torque sensor of
		the rheometer. A high-speed camera is used to capture the
		dispersed oil droplets.
		(b) Typical snapshots of the emulsion for various $Re$ at a given oil volume fraction $\phi=1\%$. From left to right are the cases of $Re=7.81\times10^3$, $1.04\times10^4$ and $2.08\times10^4$. All three cases are recorded with a high-speed
		camera connected with a long distance microscope. 
	}
	\label{Fig1_1} 
\end{figure*}

The emulsion in our study consists of oil and an aqueous
ethanol-water mixture. The silicone oil (Shin-Etsu KF-96L-2cSt) used in this study has a viscosity of
$\nu_{o}=2.1 \times10^{-6}\rm~m^{2}/s$ and a density of
$\rho_{o}=866\rm~kg/m^{3}$. The aqueous ethanol-water mixture
($\nu_{w}=2.4\times10^{-6}\rm~m^{2}/s$, $\rho_{w}=860\rm~kg/m^{3}$) is
prepared with $75\%$ ethanol and $25\%$ water in volume to match the
density of the oil. The viscosity values are measured with a hybrid rheometer type of TA DHR-1 at a temperature of $T=22
^\circ$C. The silicone oil and the ethanol-water solution are immiscible. In all experiments, no surfactant is added. In the current work, the oil volume fraction is kept at $\phi \le 40\%$, the dispersed phase always being the oil
droplets.  Though the two liquids are almost density matched, the
emulsion still tends to separate after they are mixed without adding
surfactants and in absence of an external stirring. Considering the meta-stability of
the mixture of oil and ethanol-water, a Taylor-Couette turbulent flow
is used to stir the emulsion towards a dynamical equilibrium
state. Basically, we input energy via the rotation of the inner cylinder to maintain the system in a turbulent emulsified state.
If the forcing is stopped, the emulsions coarsen until it is fully destroyed with the two immersible fluids fully separated.

The experimental setup is shown in figure \ref{Fig1_1}(a). A
Taylor-Couette system is constructed from a rheometer (Discovery
Hybrid Rheometer, TA Instruments).  The system has an inner cylinder
radius of $r_{i}=25\rm~mm$, an outer cylinder radius of
$r_{o}=35\rm~mm$, a gap $d=10\rm~mm$ and a height of
$L=75\rm~mm$. These give a radius ratio of $\eta=r_{i}/r_{o}=0.71$ and
an aspect ratio of $\Gamma=L/d=7.5$. The inner cylinder is made of
aluminum, and the outer one is made of glass. The inner cylinder is
connected to the torque sensor of the rheometer (with an accuracy
of $0.1\rm~nN\cdot m$). 
\textcolor{black}{
	The control parameter of the Taylor-Couette flow is
	the Reynolds number defined as: 
	\begin{equation}
	Re=\omega_{i}r_id/\nu,
	\end{equation}
	and the response parameter is the dimensionless torque given by: 
	\begin{equation}
	G=\frac{\tau}{2\pi L\rho\nu^{2}},
	\end{equation}
	where $\tau$ denotes the torque that is required to maintain the inner cylinder rotating at a constant angular velocity $\omega_i$, and $\nu$ is the viscosity of the emulsion.}
By rotating the inner cylinder with an imposed angular velocity, the emulsion is formed when it achieves a dynamically statistical equilibrium state, characterized by a detected balance between the break-up and the coalescence of the oil droplets dispersed in the ethanol-water continuous solution. After that the system has reached a statistically stable state, the direct measurements of the time series of the torque are recorded with the torque sensor. From this, we compute a time-averaged value of the torque. Experiments are conducted for different oil fractions, $\phi$, and angular velocities, $\omega_{i}$. The temperature of the emulsion system is maintained at $T=22\pm1^\circ$C by controlling the time duration of
each experiment, and the effect of temperature on the physical parameters
(viscosity, interfacial tension) can be neglected. 
A high-speed camera Photron (NOVA S12) is used to record the dispersed oil droplets in the ethanol-water solution. 
\textcolor{black}{Considering that the size of the droplets (about $40-500~\si{\micro\meter}$) and that of the measurement window ($3\rm~mm$) are both much smaller than the diameter of the outer glass container ($80\rm~mm$), the distortion due to the curvature can be neglected.
	To ensure achieving enough statistics, the average droplet size is computed from $\mathcal{O}(10^{3})$ samples, for all experiments.} All experiments are performed at room temperature, $T=22\pm1^\circ$C, and under atmospheric pressure conditions.

\section{Results and discussion}

\subsection{Statistical properties of droplets at a low volume fraction}

The size distribution of dispersed droplets is an important statistical parameter, as it characterizes the microscopic structure of the turbulent emulsion, which closely links to the macroscopic rheological properties and the global transport properties of the fluid system. At a low volume fraction, the droplet sizes in the turbulent emulsion eventually show a statistically stationary distribution for the current experiments under stationary stirring conditions.

Figure \ref{Fig1_1}(b) shows some typical snapshots of the emulsion for
three different Reynolds numbers $Re=\omega_{i}r_id/\nu$. Since in these cases the volume fraction of the oil phase is very low ($\phi=1\%$), the viscosity of the emulsion is approximately equal to that of the continuous phase, i.e., $\nu=\nu_{w}$.
It is expected that the average droplet size of the emulsion at a higher $Re$ will be smaller than that at a lower $Re$. The reason is that the higher average shear strength, in the cases of larger $Re$, promotes the breakup of oil droplets.  

\begin{figure*}
	\centering
	\includegraphics[width = 1\textwidth]{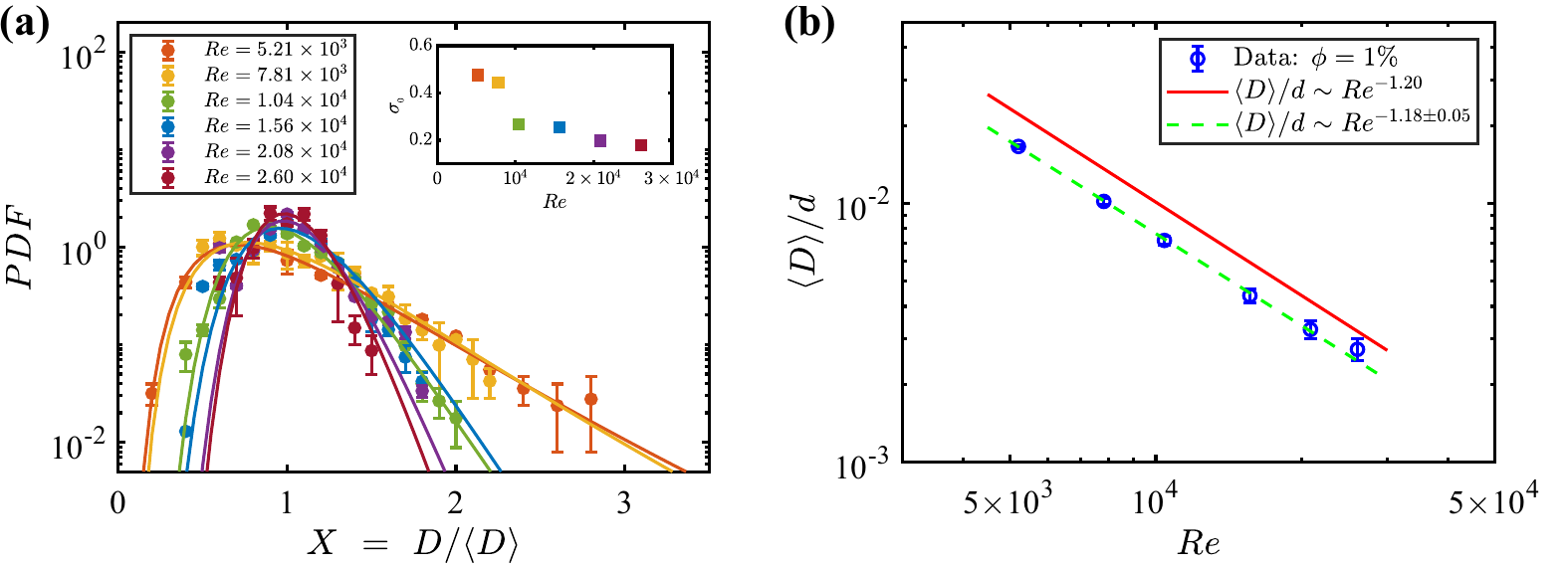}
	\caption{ 
		\textcolor{black}{
			(a) The probability density function (PDF) of the droplet diameter, with respect to the average diameter, for various Reynolds numbers, $Re$. The solid lines denote the fitting results with a log-normal distribution function. The statistics are based on
			$\mathcal{O}(10^{3})$ droplet samples for each $Re$ value. The statistical error bars are shown for all $Re$ cases. The fitted values of the standard deviation $\sigma_{0}$ as a function of $Re$ are shown in the inset.
		}
		(b) The average droplet diameter normalized
		by the gap width as a function of the Reynolds number.
		The blue circles are the data of the droplet diameter
		for $\phi=1\%$ measured in experiments, and the error bars
		are based on the errors of the edge detection.
		The red solid line denotes the power-law dependence based on the Hinze relation using the local energy dissipation rate in the bulk (Eq.~\ref{eq:Hinze_2}). The green
		dashed line represents the weighted fit of the experimental
		data, and the weight is based on the relative error of each
		data point.
	}
	\label{Fig1_2} 
\end{figure*}

Droplets interface are extracted from the recorded images, at various Reynolds
numbers, and the diameter of all the detected droplets is calculated and
normalized with the average droplet diameter, for each $Re$ cases, as $X
= D/\left<D\right>$. The distribution of the number of droplets of
size $X$, as a function of $X$, is computed as the probability
density function (PDF) of the droplet size and shown in figure \ref{Fig1_2}(a),
for various Reynolds numbers.
\textcolor{black}{
	It is clearly shown (the solid lines in figure \ref{Fig1_2}(a)) that the experimental results at all Reynolds numbers can be well described with the log-normal distribution
	\begin{equation}
	P(X) = \frac{a}{X\sigma_0\sqrt{2\pi}}\text{exp}\left[-\frac{\left[ \text{log}(X) - \text{log}(X_0) \right] ^2}{2\sigma_0^2}\right],
	\end{equation}
	where $a$, $X_0$, and $\sigma_0$ are fitting parameters. These log-normal distributions suggest that fragmentation is the primary process for droplet generation in the current system. Similar fragmentation processes are also observed in other systems \citep{villermaux2007fragmentation}, including plume formation in Rayleigh-B\'enard turbulence \citep{Zhou07,Ahlers12} among others. In addition, it is found that the fitted value of the standard deviation $\sigma_0$ decreases monotonously with increasing $Re$ (see the inset of figure~\ref{Fig1_2}(a)). This means that the distribution of droplet size becomes narrower at higher $Re$, as clearly shown in figure \ref{Fig1_2}(a).
	Some additional analyses of the distribution of droplet size are provided in appendix \ref{appC} using the gamma distribution function, which is found to be a good description of the droplet break-up during the atomization process~\citep{villermaux2007fragmentation}.
}
The next question is what sets the droplet size in the typical size in the fragmentation process leading to the droplet formation.

In 1955, Hinze proposed that the maximum stable droplet diameter in a
homogeneous and isotropic turbulent flow is given by $D =
C(\rho_w/\gamma)^{-3/5}\varepsilon^{-2/5}$, where, $\rho_w$ is
the density of the continuous phase (the ethanol-water solution in the present case), $\gamma$ is the surface tension between the two phases, $\varepsilon$
is the energy dissipation rate, and the coefficient $C = 0.725$ was
obtained by Hinze through fitting with the experimental data available at that time \citep{hinze1955fundamentals}. The argument of Hinze
applies to dilute distribution of droplets that occasionally coalesce
due to collisions and breakup due to turbulent stresses. A key element of Hinze's argument consists in assuming that droplets do not produce a significant feedback on the turbulent flow, whose statistics is the one of homogeneous and isotropic turbulence. Many studies show that the average droplet size and the maximum size are proportional in turbulent emulsions \citep{lemenand2003droplets,boxall2012droplet}. Considering the maximum droplet diameter in turbulent emulsions is usually unstable due to the breakup and occasional coalescence, the average droplet diameter can be used as the indicator of the droplet size in the Hinze relation \citep{perlekar2012droplet}. 

We notice that the distribution of the energy
dissipation rate in Taylor-Couette turbulence is inhomogeneous, i.e., the dissipation in the bulk is much smaller than that in the boundary layers.  As the volume of the bulk is much larger than that of the boundary layers in the
current parameter regime \citep{grossmann2016high}, droplets are
expected to mainly distribute in the bulk, where the flow is found
to be nearly homogeneous and isotropic \citep{ezeta18jfm}.  The local
energy dissipation rate in the bulk can be expressed as
$\varepsilon_l\sim u_{_T}^{3}/\ell$, where $u_{_T}$ and $\ell$
denote the typical velocity fluctuation and the characteristic
length scale of the flow. The typical velocity fluctuation $u_{_T}$
can be approximated as A$\omega_{i}r_{i}$ in the current Taylor-Couette
turbulent flow \citep{vanGils2012} with an almost constant prefactor A (order of
0.1).  As we know that the Reynolds number can
be expressed as $Re=\omega_{i}r_{i}d/\nu$, then the typical velocity
fluctuation can be expressed as $u_T \sim \omega_{i}r_{i} \sim Re\cdot
\nu /d$. Plugging this velocity estimation into the expression for
the energy dissipation above, we obtain $\varepsilon_l \sim
u_{_T}^{3}/\ell \sim Re^3\nu^3/d^4$ by assuming $\ell \sim
d$, and this scaling dependence is also in good agreement with the recent measurement of the local energy dissipation rate in the bulk of Taylor-Couette turbulence \citep{ezeta18jfm}. Inserting this local energy dissipation expression into Hinze's relation, one obtains
\begin{equation}
\left<D\right>/d \sim C/d\left(\frac{\rho_w}{\gamma}\right)^{-3/5}\varepsilon_l^{-2/5}
\sim Re^{-6/5},
\label{eq:Hinze_2}
\end{equation}
suggesting that the average droplet diameter has a power-law dependence on $Re$ with an effective power-law exponent of $-$1.20
(Eq.~\ref{eq:Hinze_2}). We compare the dependence of the normalized droplet size on $Re$ from the experiments and the model in figure \ref{Fig1_2}(b). The best fit of the experimental data gives a scaling exponent of $-1.18 \pm 0.05$. We find that the scaling dependence based on the local energy dissipation rate in the bulk (red solid line) agrees well with the experimental data.
\textcolor{black}{
	The results show that the scaling dependence of the droplet size on $Re$ could be connected to turbulent fluctuations in the bulk of the system. The discussion above is a simple analysis based on the scaling law, more in-depth and quantitative understanding of the droplet formation in a turbulent (Taylor-Couette) emulsion flow deserves further studies in the future.
}

\subsection{Effective viscosity and shear thinning effects}
\begin{figure*}
	\centering
	\includegraphics[width = 1\textwidth]{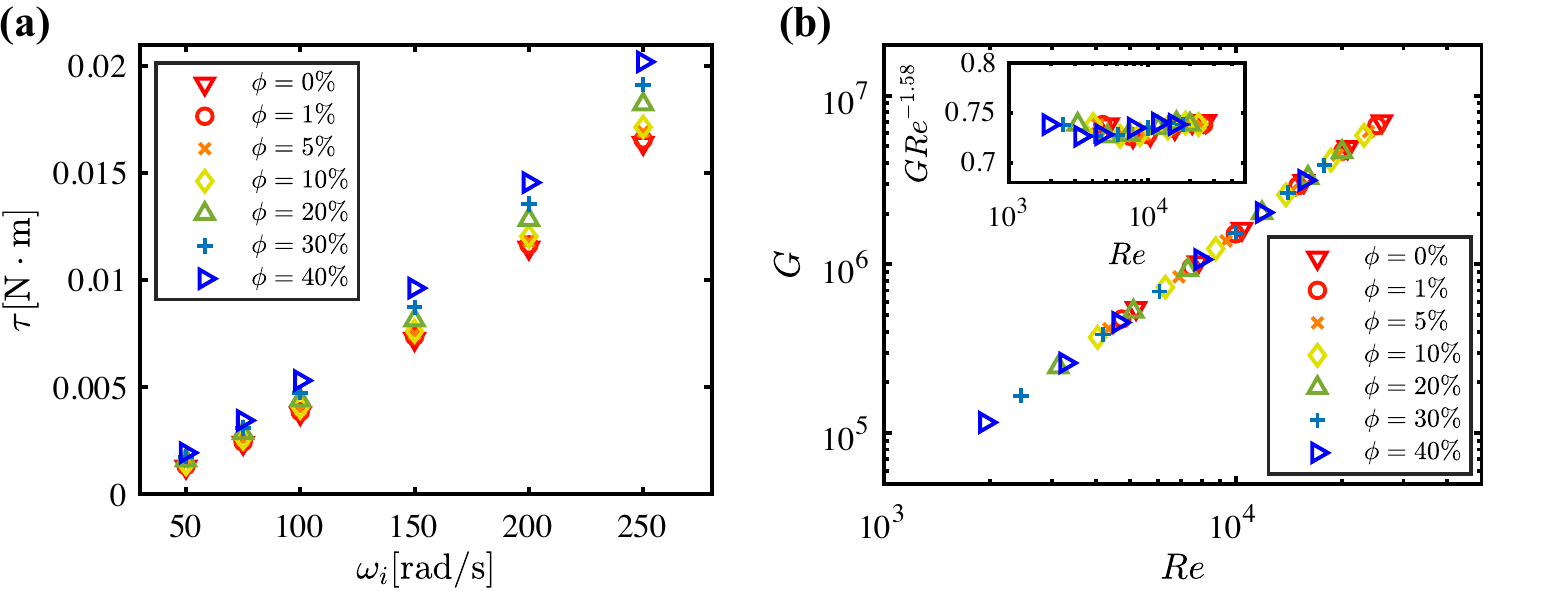}
	
	\caption{
		(a) Torque measurements. The torque that is
		required to maintain the inner cylinder at a constant angular
		velocity, $\omega_{i}$, is measured for the emulsion systems with
		oil volume fractions corresponding to $\phi=0\%, 1\%, 5\%, 10\%,
		20\%, 30\%$, and $40\%$. 
		As an estimate for the error on the torque measurements, we use its standard deviation and find that it is smaller than $0.8\%$ for all volume fractions (see appendix~\ref{appA} for details). The error results are therefore smaller than the symbol size. 
		(b) The dependence between the
		dimensionless torque, $G$, and the Reynolds number, $Re$, by using the effective viscosity. All
		these sets of data at the various oil volume fractions collapse in
		a master curve, and the error is less than $1\%$. The inset shows the dimensionless torque compensated with $Re^{-1.58}$.}
	\label{Fig2} 
\end{figure*}

The torque of the inner cylinder is directly measured by the rheometer
sensor for different oil volume fractions, $\phi$, and angular
velocities, $\omega_{i}$, as shown in figure \ref{Fig2}(a), which shows
that the faster the inner cylinder rotates, the larger the torque is
needed to maintain the selected angular velocity. The torque becomes
larger when the oil volume fraction is increased at a given angular
velocity, indicating that the oil additive will bring an obvious
change to the rheological property of the emulsion system. Combined with the flow properties of Taylor-Couette turbulence at various Reynolds numbers
\citep{grossmann2016high,van2011torque,van2011twente,ostilla2014boundary,huisman2014multiple}, we can calculate the effective viscosity of emulsions in these dynamical equilibrium states. We use the same method which was recently proposed for the viscosity measurements in a very high Reynolds number Taylor-Couette flow \citep{Bakhuis2020}. 

An effective power-law dependence between $G$ and $Re$ can be obtained as $G \propto Re^{\beta}$ for the Taylor-Couette turbulent flow, and the power-law exponent $\beta$ depends on the Reynolds number regime
\citep{grossmann2016high}. Here we assume that the power-law dependence $G \propto Re^{\beta}$ can still be applied to the two immiscible
liquids in our Taylor-Couette turbulent flow.  As a reference
case, this relation can be determined by using the results of the pure
ethanol-water mixture ($\phi=0\%$) with a known viscosity. When we
plot together all data for the various oil fractions in a $G$-$Re$ plot,
and collapse them on a master curve with an effective exponent of
$\beta$ = 1.58 (figure \ref{Fig2}(b)), the effective viscosity is a
fitting parameter for each cases. To demonstrate the
quality of the overlap of the different data sets, all data are
compensated by $Re^{1.58}$ (inset of figure \ref{Fig2}(b)), which
clearly shows that the effective power-law dependence works very well. Remarkably, the power-law dependence $G \propto Re^{1.58}$ for single-phase Taylor-Couette flows still works well for the present two-phase emulsion flows.
By using the effective power-law exponent of $\beta$ = 1.58 between $G$ and $Re$, we can calculate the effective viscosity of the emulsion at various $\omega_{i}$ and $\phi$ with an expression of
$\nu_{eff}=\nu_{w}(\tau/\tau_{w})^{2.38}$; here $\nu_{w}$, $\tau_{w}$ and $\tau$ denote the viscosity
of the ethanol-water solution, the measured torque of ethanol-water
solution and that of the emulsion, respectively. It should be noted
that $\nu_{w}$ is known, but $\nu_{eff}$, $\tau_{w}$ and $\tau$ are
dependent on the experimental settings, i.e. $\phi$ and $\omega_{i}$.
\textcolor{black}{
	The detailed calculation of the effective viscosity is documented in appendix \ref{appB}.
}

\begin{figure*}
	\centering
	\includegraphics[width = 1\textwidth]{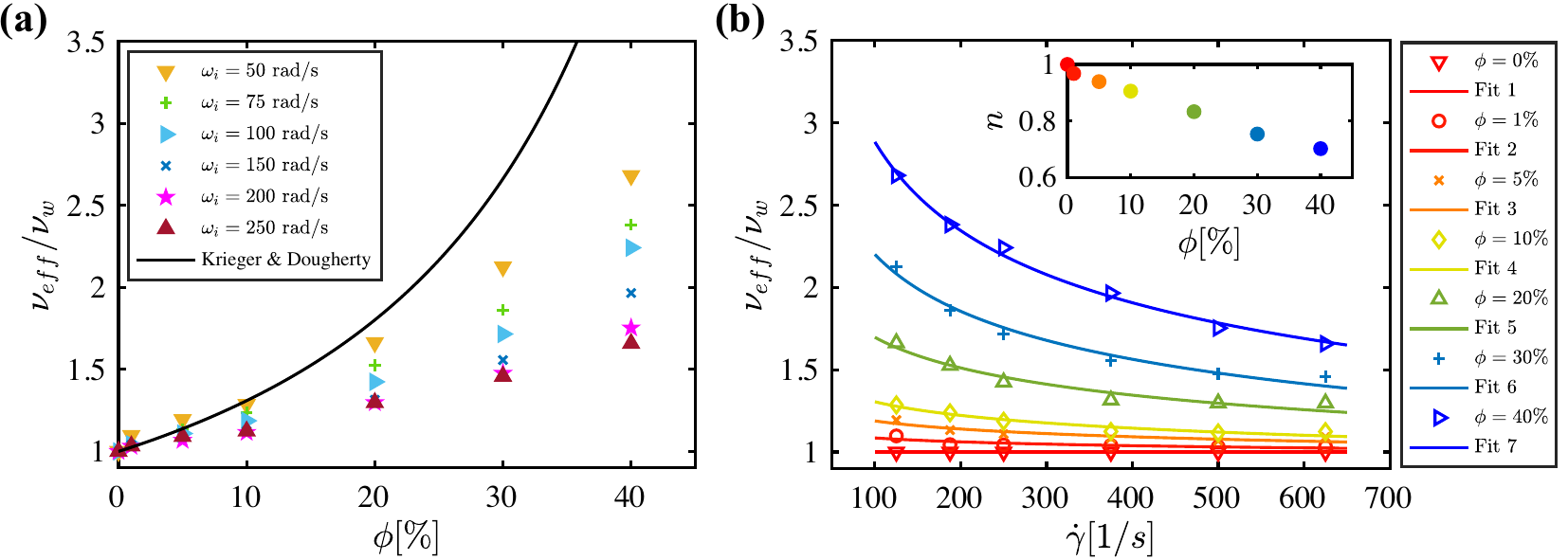}
	
	\caption{(a) The effective viscosity normalized by the viscosity of
		the ethanol-water mixture, $\nu_{eff}/\nu_{w}$, as a function of
		the volume fraction of the dispersed oil phase $\phi$ at various angular velocity $\omega_{i}$. The solid line
		denotes the effective viscosity model for solid particle suspensions by Krieger $\&$ Dougherty \citep{krieger1959mechanism}. The calculation of the effective viscosity
		is based on the torque measurements. The relative standard
		deviation is less than $2.5\%$, so the errors bars are smaller
		than the symbol size.  (b) The effective viscosity of the
		emulsion versus the characteristic shear rate $\dot{\gamma}$ of
		the flow. The data for the different oil volume fractions, $\phi$,
		are denoted by the hollow symbols with various colors in the
		legend. The solid lines show the fitting results (Fit 1-7) using
		Herschel-Bulkley model \citep{HB-model} for various volume
		fractions. The inset shows the power-law index $n$ as a function
		of the volume fraction $\phi$. }
	\label{Fig3} 
\end{figure*}


To understand the effect of oil addition on the rheology of the
emulsion in turbulent shear flows, we systematically vary two
parameters of the system, i.e., the oil volume fraction, $\phi$, and
the angular velocity of the inner cylinder, $\omega_{i}$. The effective viscosity of emulsions, as a function of $\phi$ and for various $\omega_{i}$, is reported in figure \ref{Fig3}(a), where all data are normalized by the viscosity of the ethanol-water solution, $\nu_{w}$. Obviously, the effective viscosity of the emulsion increases
with increasing the oil volume fraction, $\phi$, for all $\omega_{i}$
cases. While the effective viscosity has a weak dependence on $\phi$
in the dilute regime (e.g. for $\phi<5\%$), it displays a stronger
dependence for larger $\phi$.  The hydrodynamic or contact
interactions between oil droplets for larger $\phi$ are expected to
yield an increasing viscous contribution, somehow similar to what
observed for the case of dispersion of hard spheres in suspensions
\citep{guazzelli2018rheology}. The
relation between the effective viscosity and the volume fraction of
dispersed solid particles in particle-fluid suspensions is also plotted in figure \ref{Fig3}(a) for comparison. Strictly speaking, we find that the effective viscosities of the emulsions, at all $\omega_{i}$, are smaller than that of the dependence proposed by \cite{krieger1959mechanism}. Here it
needs to be emphasized that this model was developed for
suspensions of monodispersed hard spheres in fluids in the viscous
regime. This disagreement in the
viscosity can be due to the different nature of the dispersed phases:
the dispersed solid particles have a fixed, undeformable shape, while
the dispersed oil droplets can deform; the solid particles have a fixed
size, while the droplets can dynamically coalesce and break up under the
flow. The dynamics of the dispersed droplets in emulsions are therefore much
richer than that of the solid particles in suspensions.

Furthermore, the effective viscosity is found to decrease with
increasing $\omega_{i}$ for a given volume fraction, $\phi$, as indicated in
figure \ref{Fig3}(a). In other words, the turbulent emulsion shows a continued shear
thinning behavior. To reveal this effect better, we plot the effective
viscosity as a function of shear rate in figure \ref{Fig3}(b), here the
shear rate is defined as $\dot{\gamma}=\omega_{i}r_{i}/d$. Though the
Taylor-Couette flow is not a planar shear flow, the shear rate
$\dot{\gamma}$ can still represent well the effective shear strength of
the system. When the volume fraction of oil is $\phi=0\%$ (i.e. pure
ethanol-water), the system is a single-phase flow state and, as
expected, the effective viscosity does not change with the shear rate, 
$\dot{\gamma}$. With the addition of the oil phase, the effective
viscosity of the emulsion decreases with increasing $\dot{\gamma}$,
and this effect is more pronounced for high
volume fractions, as shown in figure \ref{Fig3}(b).
\textcolor{black}{
	This shear thinning behavior is similar to what was found in a suspension of deformable microgel particles under steady shear flow~\citep{adams2004influence}.
}

To quantify the shear thinning effect of the turbulent emulsion, we
compare our data with the Herschel-Bulkley model \citep{HB-model}:
\begin{equation}
\mu_{eff}=k_{0}\dot{\gamma}^{n-1}+\tau_{0}\dot{\gamma}^{-1},
\end{equation} 
where $\mu_{eff}$ is the effective dynamic viscosity, $k_{0}$ and $n$
represent the consistency and the flow index, respectively,
and $\tau_{0}$ is the yield shear stress. As the system is far from the jamming state, the yield shear stress is expected to be zero in the current case
($\tau_{0}=0$). Consequently, the Herschel-Bulkley model can then be
simplified as $\nu_{eff}/\nu_{w}=K\dot{\gamma}^{n-1}$. The fitting
results using the Herschel-Bulkley model for various volume fractions
are also shown in figure \ref{Fig3}(b). As expected, the flow index is
around 1 at very low volume fractions, suggesting that the fluid
behaves like a Newtonian fluid. The flow index, $n$, monotonically
decreases with an increasing volume fraction of dispersed phase (inset of
figure \ref{Fig3}(b)), indicating a more pronounced shear thinning effect
for the emulsions with high oil volume fractions. The agreement between the
experimental data and the Herschel-Bulkley model indicates
that the shear thinning effect can be well described by this classical
non-Newtonian model, opening an important avenue for the description
of the effective viscosity of the turbulent emulsion systems.

\section{Conclusions}

Turbulent emulsions are complex physical systems coupling macro- and
micro-scales.  In this work, we investigated the dynamics of the emulsions
of oil droplets dispersed in an ethanol-water solution without surfactant additive in a turbulent shear flow. Firstly, we find that the PDF of the droplet sizes follows a log-normal distribution, suggesting a fragmentation process in the droplet generation process. 
\textcolor{black}{
	By comparing the droplet size for various Reynolds numbers for the system at a low volume fraction of $1 \%$ with Hinze theory, we find that the scaling dependence of the droplet size on Reynolds number can be connected to the turbulent fluctuations in the bulk of the system.
}

The effective viscosity of the emulsion is found to increase with
increasing the oil volume fraction, but the increasing trend is weaker
than the one reported for solid particle suspensions. This difference
is associated with the different nature (deformability and size distribution) of the dispersed phase in the fluid-fluid emulsions.
Additionally, we find that the effective viscosity of the emulsions
decreases at increasing the shear rate, displaying a shearing thinning
behavior that can be quantitatively described using the
classical Herschel-Bulkley model via a dependency of the flow index on
the volume fraction. 
The shear thinning effect of the turbulent emulsion has many potential
applications, such as drag reduction of multi-component liquid systems
in turbulent states. The current findings have important implications for extending the knowledge on turbulence and low-Reynolds-number emulsion flows to turbulent emulsion flows.

\vspace{-4 mm}

\section*{Acknowledgements}
We thank Frederic Risso, Detlef Lohse, Sander Huisman, and Thomas van Vuren for insightful suggestions and discussions, and thank Huiling Duan, Pengyu Lyu, Baorui Xu, and Yaolei Xiang for the help with the experimental setup. This work is financially supported by the Natural Science Foundation of China under Grant No.~11988102, 11861131005, 91852202 and 11672156, and Tsinghua University Initiative Scientific
Research Program (20193080058).

\vspace{-4 mm}
\section*{Declaration of interests}
The authors report no conflict of interest.

\appendix
\section{Experiments}\label{appA}
\subsection{Liquids used in the current study}
We use silicone oil (dispersed phase) and ethanol-water mixture (continuous phase) in current experiments. The silicone oil and ethanol-water solution are immiscible. The density of silicone oil type of Shin-Etsu KF-96L-2cSt is $\rho_{o}=866\rm~kg/m^{3}$. We use an aqueous mixture of deionized water and ethanol as the second liquid. The volume fraction of water is $25\%$. The density of the ethanol-water mixture is $\rho_{w}=860\rm~kg/m^{3}$, which is very close to that of silicone oil. The density match of these two kinds of liquids can eliminate the effect of centrifugal force on liquid distribution. Both ethanol-water mixture and silicone oil are transparent, which facilitates the imaging of emulsions. As the refractive indices of these two kinds of liquids are different, we can distinguish the oil droplets from the background of ethanol-water.

\begin{figure}\centering
	\includegraphics[width = 0.6\textwidth]{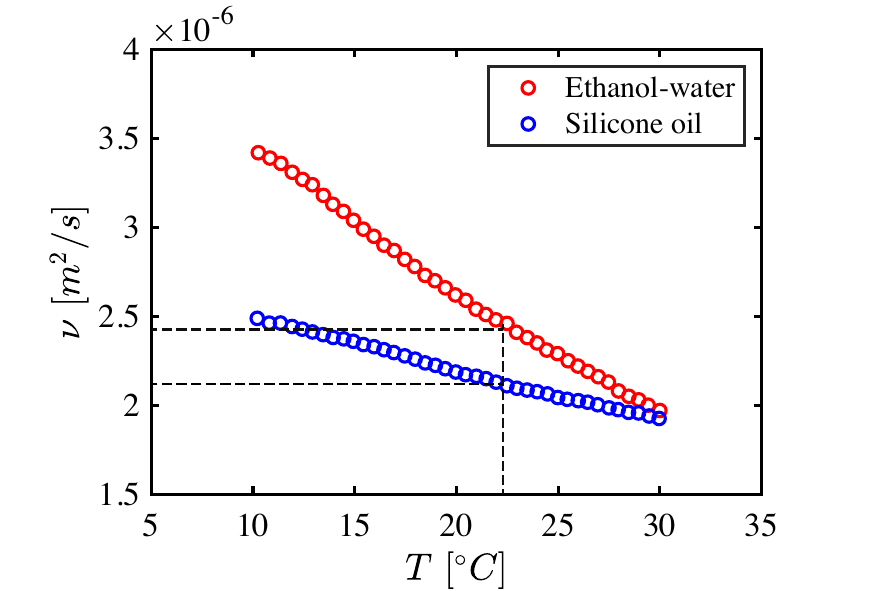}
	\caption{
		The kinematic viscosity $\nu$ as a function of temperature $T$. The red circles denote measured viscosity of ethanol-water, and the blue circles denote that of silicone oil. For the temperature of experiments $T=22^\circ$C, the viscosity of ethanol-water is $\nu_{w}=2.4\times10^{-6}\rm~m^{2}/s$, while that of silicone oil is $\nu_{o}=2.1\times10^{-6}\rm~m^{2}/s$. 
	}
	\label{v_T}
\end{figure}

The viscosity is measured by using a hybrid rheometer type of TA DHR-1 (Discovery Hybrid Rheometer, TA Instruments). We equip the rheometer with a parallel plate, which is appropriate for measurements of low-viscosity liquids in the current study. The Peltier plate steel under the measured liquids provides temperature control and measurement with an accuracy of $\pm0.1^\circ$C. The plots of kinematic viscosity $\nu$ versus temperature $T$ for these two kinds of liquids are shown in figure~\ref{v_T}. The viscosity of ethanol-water is larger than that of silicone oil in the measured temperature range of $10^\circ$C to $30^\circ$C. At the experimental temperature $T=22^\circ$C, we found the viscosity of ethanol-water is $\nu_{w}=2.4\times10^{-6}\rm~m^{2}/s$, which is close to that of silicone oil $\nu_{o}=2.1\times10^{-6}\rm~m^{2}/s$. 
The interfacial tension between the dispersed phase and continuous phase is an important parameter in emulsions, which closely links to the break-up and coalescence of droplets. We measure the interfacial tension between the two kinds of liquids (ethanol-water and silicone oil) used in the current experiments with the pendant drop method. The type of measurement instrument is SCA20. The interfacial tension is calculated by using characteristic parameters of the drop profiles and density difference of the liquids.
We perform 6 measurements and use the average value as the final result of interfacial tension $\gamma=4.53\rm~mN/m$. All measurements are conducted at a temperature of $T=22^\circ$C.

\begin{figure}\centering
	\includegraphics[width = 0.9\textwidth]{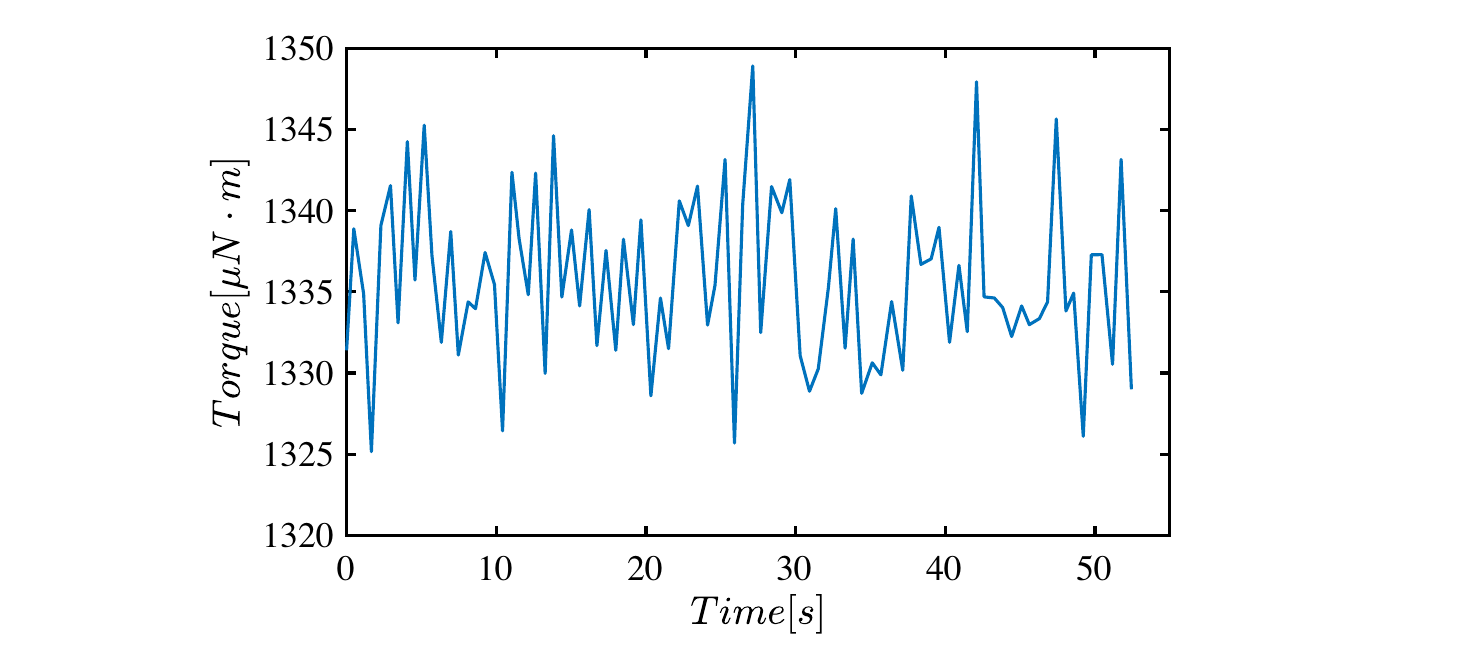}
	\caption{
		A typical result of time series of the torque measurements for emulsion system. The oil volume fraction is $\phi=1\%$, and the Reynolds number is $Re=5.21\times10^{3}$ in this case. The standard deviation of the time series of torque is 15.47 $\si{\micro}\rm N\cdot m$, which is much smaller than the averaged torque value.
	}
	\label{torque}
\end{figure}

\begin{figure}\centering
	\includegraphics[width = 1\textwidth]{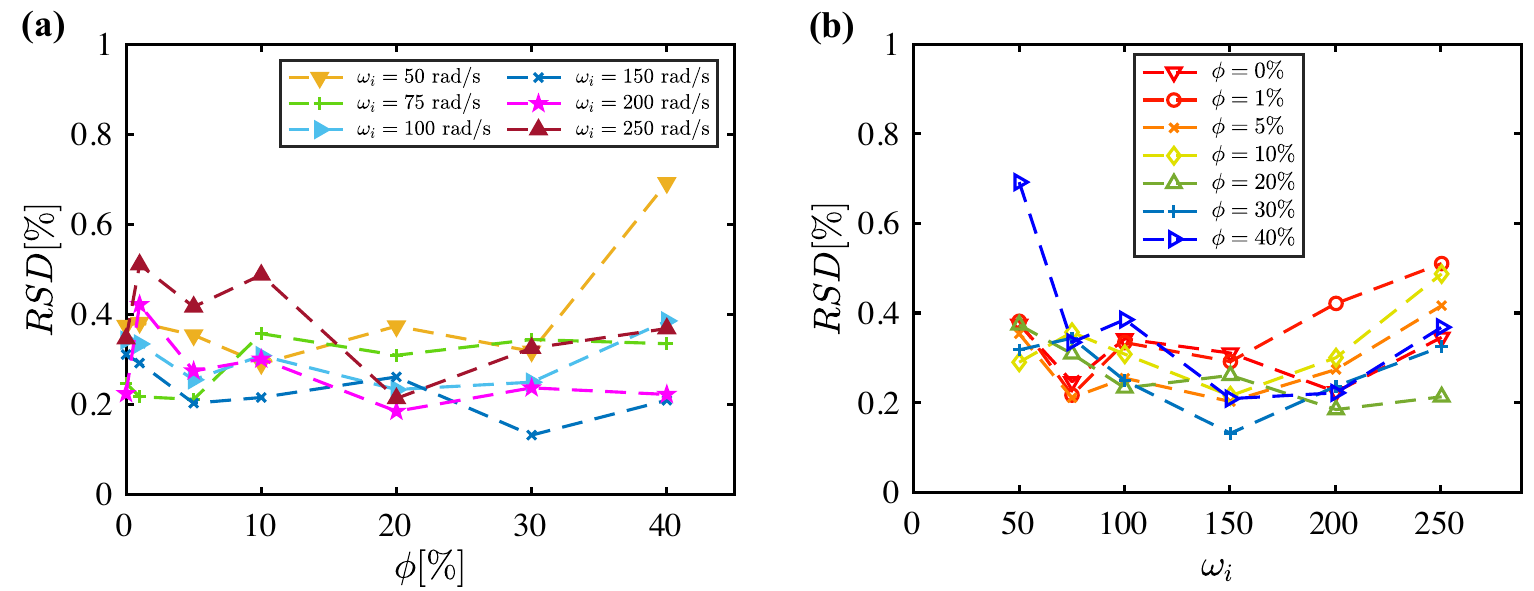}
	\caption{
		(a) The relative standard deviation (RSD) of torque time series as a function of the volume fraction for various angular velocities.
		(b) The relative standard deviation (RSD) of torque time series as a function of the angular velocity for various volume fractions.
	}
	\label{torque_RSD}
\end{figure}
\subsection{Torque measurement}
The torque is a response parameter of the emulsion system in the current study. The torque is directly measured by the rheometer through the shaft connected to the inner rotating cylinder with high accuracy up to $0.1\rm~nN\cdot m$. For each experiment, we set the angular velocity $\omega_{i}$ of the inner cylinder as a constant value. After the system reaches a statistically stable state, the direct measurements of time series of torque are recorded. The typical time series of torque measurements are shown in figure~\ref{torque}. The standard deviation of the torque time series is 15.47 $\si{\micro}\rm N\cdot m$, which is much smaller than the torque value and consequently fulfills the requirement of the torque measurement. To show the quality of torque measurements, we calculate the relative standard deviation (RSD) for all cases in the current study, as shown in figure~\ref{torque_RSD} (a)(b). We find that all values of RSD are smaller than $0.8\%$, indicating that the torque measurements are reliable. The results show that the RSD doesn't change with oil volume fraction $\phi$ and angular velocity $\omega_{i}$.

\begin{figure}\centering
	\includegraphics[width = 1\textwidth]{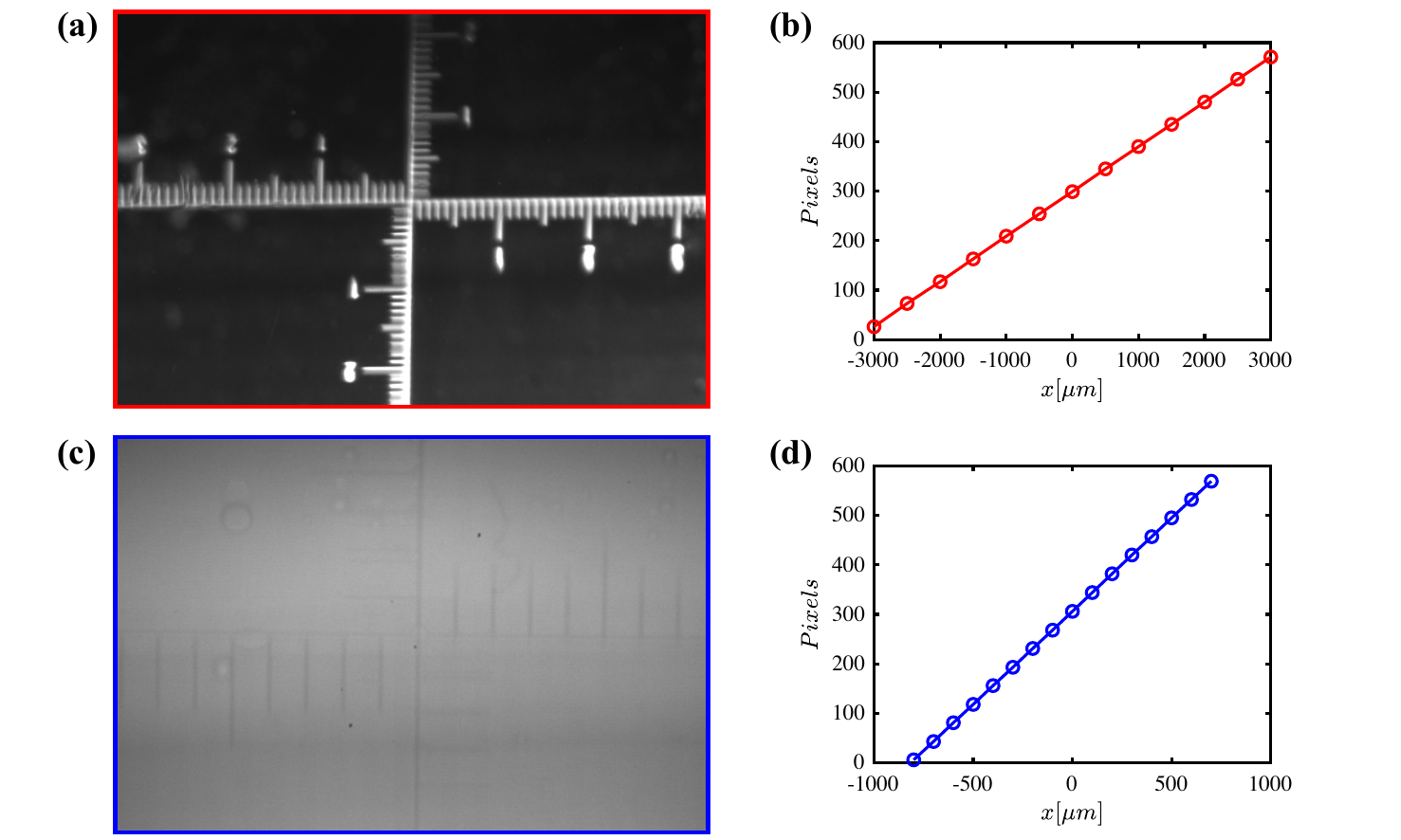}
	\caption{
		The results of length calibration for two sets of lens used in the current study. (a) The image of length calibration for Set 1. Each smallest tick interval is $100~\si{\micro\meter}$. (b) The calibration results of Set 1. The x-axis corresponds to the ticks in (a), and the y-axis corresponds to the pixel distance from the left border in (a). (c) The image of length calibration for Set 2. Each smallest tick interval is $100~\si{\micro\meter}$. (d) The calibration results of Set 2. The x-axis corresponds to the ticks in (c), and the y-axis corresponds to the pixel distance from the left edge in (c). 
	}
	\label{calibration}
\end{figure}

\subsection{Imaging of the dispersed drops}
The statistical properties of dispersed oil droplets in emulsions are important parameters in the current study. We use a high-speed camera to capture the drops, which are constantly moving fast along with the flow in turbulent states. Two sets of camera lenses are used. One is a Nikon 105mm f/2.8G macro lens with an extension tube that gives about $2\times$ magnification ratio (Set 1). This set of lenses is used for a low $Re$ case ($Re = 5.21 \times 10^3$), in which the drop size is in the range of about $40-500~\si{\micro\meter}$. The light source is two front lamps, and the reflected light from the surface of the inner cylinder is used for imaging. For the experiments at higher $Re$ ($Re>5.21\times10^3$), another set of NAVITAR microscopic lens coupled with a $5\times$ objectives type (Mitutoyo M Plan Apo.) is connected to the high-speed camera to resolve the very small oil droplets in turbulent Taylor-Couette flows (Set 2). For this set of lenses, the light source is coaxial with the microscope so that we can obtain a better view in the small observation area. The axes of the lens are at about half the height of the system so that we can reduce the edge effects from top and bottom. 

To reduce the effect of curvature, both these two sets of the lens are focused on the central area of the TC system. For both these two sets of the lens, we perform the length calibration before experiments. The typical results of length calibration are shown in figure~\ref{calibration}. The length of the images is 600 pixels, and the height is 400 pixels. figure~\ref{calibration}(a, c) show the calibration images for Set 1 and Set 2, respectively. Each smallest tick interval is $100~\si{\micro\meter}$. We plot the pixel distance as the function of the tick distance in figure~\ref{calibration}(b, d). The linearity of the data indicates that the effect of curvature can be safely neglected in the current measurements.

\section{The effective viscosity calculation}\label{appB}
First, we calculate the $Re$ and $G$ at various angular velocities $\omega_i$ for pure ethanol-water mixture ($\phi=0\%$) with a known viscosity. When we plot these data in a $G-Re$ plot, we find a scaling law as $G\sim Re^{1.58}$. Further, we can write this relation as $G=KRe^{1.58}$, where $K$ denotes a constant prefactor. If we insert the definitions of $G$ and $Re$ to this dependence, we obtain a dependence of torque $\tau$ and viscosity $\nu$ as
\begin{equation}
\tau=AK\nu^{0.42},
\end{equation}
where $A$ equals to $2\pi L\rho/(\omega_{i}r_id)^{0.42}$.
We assume that this relation is still valid for emulsion systems with various oil volume fractions and Reynolds numbers.  We write the torque and effective viscosity of the emulsion system as $\tau$ and $\nu_{eff}$ for a constant angular velocity $\omega_i$ at a volume fraction of $\phi$. For the pure ethanol-water mixture ($\phi=0\%$) system at the same angular velocity, we obtain the measured torque value $\tau_{w}$ and the viscosity $\nu_{w}$. Based on our assumption, these two systems both follow the relation given above. Because the angular velocities of these two systems are the same, the prefactor $A$ is therefore the same too. Then, we can derive the following relation:
\begin{equation}
\frac{\nu_{eff}}{\nu_{w}}=\Big(\frac{\tau}{\tau_{w}}\Big)^{2.38}.
\end{equation}
The effective viscosity of emulsion systems $\nu_{eff}$ can be obtained based on this relation.
To further verify our assumption above, we calculate the $G$ and $Re$ for various volume fractions and angular velocities by using the effective viscosity obtained for each case. When we plot together all data of various oil fractions in a $G$-$Re$ plot, we find that all data of G versus Re collapse in a master curve. The fitting results for the oil fractions of $\phi=1\%, 5\%, 10\%, 20\%, 30\%$, and $40\%$ show that all these $6$ sets of data follow the relation $G=KRe^{1.58}$ with only an error bar of $1\%$, which strongly supports the assumption above. Here we provide a new approach for the measurement of the effective viscosity of emulsions in high-Reynolds number turbulent states.

\begin{figure}\centering
	\includegraphics[width = 0.99\textwidth]{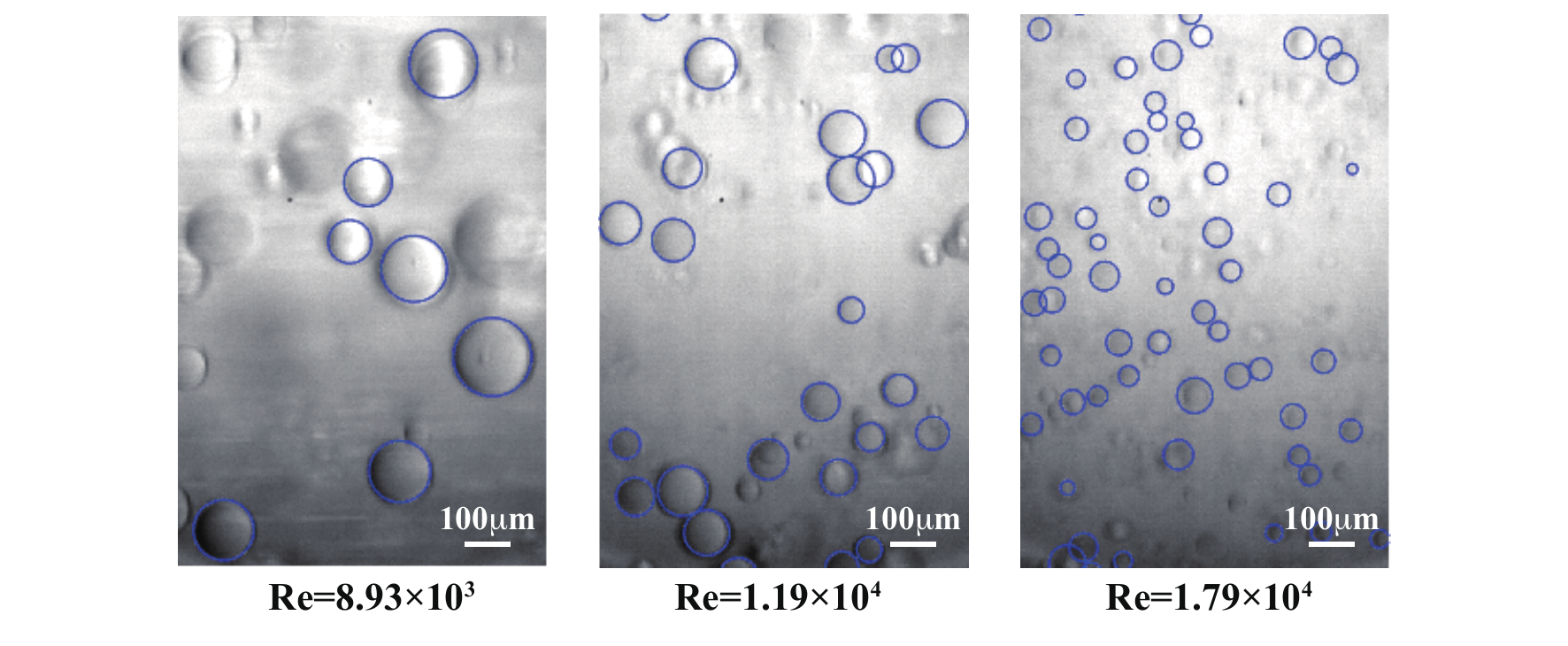}
	\caption{
		The results of drop dectection for various Reynolds numbers of $Re=7.81\times10^{3}$, $Re=1.04\times10^{4}$, and $Re=1.56\times10^{4}$. The volume fraction of oil is $\phi=1\%$ here. The blue circles in the images are the boundaries of the oil drops from the edge detection. Most of the drops in images are captured with a high fidelity.
	}
	\label{detect1}
\end{figure}

\begin{figure}\centering
	\includegraphics[width = 0.6\textwidth]{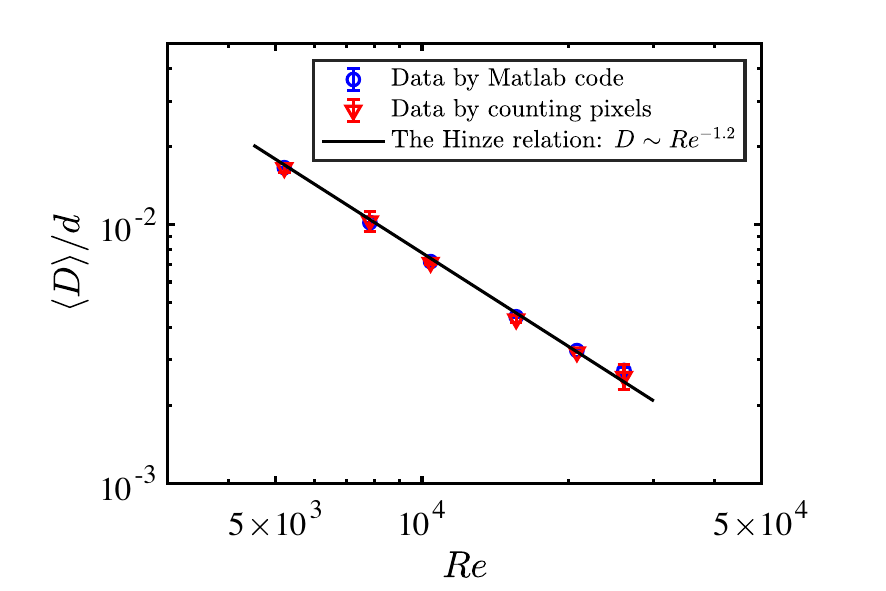}
	\caption{
		The comparison between the results of the average drop diameter detected by the Matlab code and that by manually counting pixels in the ImageJ software. The blue circles denote the data detected using the Matlab code, and the red downward triangles denote the results from the manual pixels counting. The solid black line is the fitted power-law dependence based on the Hinze relation~\cite{hinze1955fundamentals}. All data are obtained for the oil volume fraction of $\phi=1\%$.
	}
	\label{detect2}
\end{figure}

\section{The analysis of droplet size}\label{appC}
\subsection{Image processing}
The videos and images obtained in the experiments are analyzed by using the Matlab code and the ImageJ software. For better post-processing, the original images are firstly cropped and exported as the tiff-format images. The size of the clipping window is 300 pixels $\times$ 1024 pixels. At the same time, we determine the interval between every two frames based on the average speed of the droplets moving in the horizontal direction, so that the oil droplets in each image are not be counted repeatedly. 

Next, we adjust the contrast of images and detect the boundary of drops by using the Matlab code. The radii of droplets are exported as the data sets for further processing. The typical results of boundary detection for the various Reynolds numbers are shown in figure~\ref{detect1}. Most of the oil droplets in the images are well captured. Few drops are not detected, because they are out of the focal plane, inducing too blurry boundaries. Considering that we count enough droplet samples ($\mathcal{O}(10^{3})$), these undetected drops do not have much influence on the analysis of the statistical characteristics for the drops.

In order to verify the reliability of the drops detection using the Matlab code, we also use another method to calculate the droplet size. We use the ImageJ software to obtain the diameter of droplet by manually counting the pixel distance. The comparison between the results obtained by the manual counting and those by the Matlab code is shown in figure~\ref{detect2}. The differences between the two methods are very small, and they are both in a good agreement with the Hinze relation~\cite{hinze1955fundamentals}, indicating that both the results by the Matlab code and by the manual counting are reliable. The numbers of detected droplet samples at various $Re$ by the Matlab code and that by the manual counting are shown in table~\ref{Tab1}. Of course, the detection using the Matlab code provides more statistics, we therefore use the results from the Matlab detection for all cases in the main paper.
\begin{table}
	\begin{center}
		\def~{\hphantom{0}}
		\begin{tabular}{lcccccc}
			$Re$            & $5.21\times10^{3}$ & $7.81\times10^{3}$ & $1.04\times10^{4}$ & $1.56\times10^{4}$ & $2.08\times10^{4}$ & $2.60\times10^{4}$\\[3pt]
			Matlab code     & 2190 & 710 & 1643 & 2486 & 1573 & 807\\[3pt]
			Manual counting & 765 & 636 & 624 & 605 & 514 & 513\\[3pt]
		\end{tabular}
		\caption{The numbers of detected droplet samples at various $Re$ by the Matlab code and that by manual counting.}
		\label{Tab1}
	\end{center}
\end{table} 

\subsection{The distribution of droplet size}
\begin{figure}\centering
	\includegraphics[width = 0.6\textwidth]{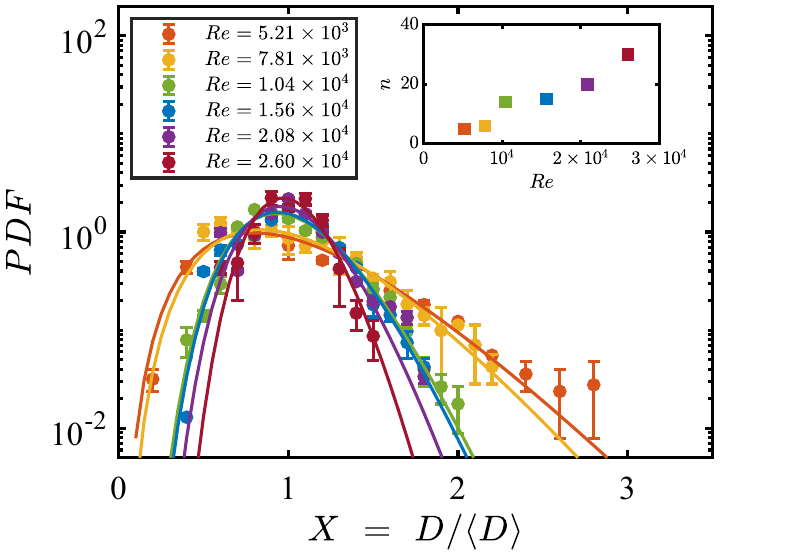}
	\caption{
		The probability density function (PDF), in log-log scale, of the droplet size for various Reynolds numbers, $Re$. The solid lines denote the fitted results using gamma distribution functions. The inset shows the fitted value of the index $n$ as a function of $Re$.
	}
	\label{PDF_gamma}
\end{figure}

\textcolor{black}{
	The distribution behaviors of droplet sizes in emulsions are found to be well described by the log-normal distribution functions. We have fitted the same data using the gamma distribution function:
	\begin{equation}
	P(X = D/\langle D \rangle) = \frac{n^n}{\Gamma(n)}X^{n-1}e^{-nX},
	\end{equation}
	where $n$ is a constant, and $\Gamma(n)$ is the gamma function.
	This function is expected to be a good description of droplet size during atomization~\citep{villermaux2007fragmentation,bremond2006atomization}.
	The results of the fit are shown in figure~\ref{PDF_gamma}. It is found that the gamma distribution function can also describe the droplet size distribution for most $Re$ cases. In addition, we also see a monotonous increase of the index $n$ with increasing $Re$, indicating that the distribution is narrowed. Indeed, we can not tell which distribution function is better for describing the distribution of the droplet size for all cases, given the current data. Thus, while we report only the results for the log-normal distribution in the main text, the results of gamma distribution are also provided here for role the of comparison. The distribution of droplet size will be studied in the future.
}

\vspace{-2 mm}


\end{document}